# Electrical Potential, Mass Transport and Velocity Distribution of Electro-osmotic Flow in a Nanochannel by Incorporating the Variation of Dielectric Constant of Aqueous Electrolyte Solution

**Rajendra Padidhapu[a], Shahnaz Bathul[b] and V.Brahmajirao[c]**
\
[a]Department of Mathematics, Flora Institute of Technology, 49/1, Khopi, Pune, Maharastra, India.
[b]Professor, Department of Mathematics, Jawaharlal Nehru Technological University, Hyderabad, India.
[c]School of Biotechnology, MGNIRSA, Hyderabad, India.
E-mail: rajendra.padidhapu@gmail.com

## Abstract

We consider a coupled system of Navier – Stokes, Maxwell – Stefan and Poisson - Boltzmann equations by incorporating the variation of dielectric constant, which governs the electro-osmotic flow in nanochannel, describing the evolution of the velocity, concentration and potential fields of dissolved constituents in an aqueous electrolyte solution. We apply the finite difference technique to solve one and two dimensional systems of these equations. The solutions give an extremely accurate prediction of the dielectric constant for a variety of salts and a wide range of concentrations.

## Introduction

Coulombs law defines the electrostatic force of interaction between charge entities. As per this law, the electrostatic behavior is governed by a very important property of the medium in which the charges are situated, and it is called electrostatic permittivity or dielectric constant. As the dielectric constant increases, the medium permits fewer electrostatic lines of force through it. Consequently, the force of interaction decreases. Dielectric medium has the property of insulating lines of force and tends to stop the flow of charge (Robinson and Stokes, 1959).The dielectric permittivity is a very important characteristic but not properly accounted in the theories so far. Starting from Debye-Hückel theory of electrolytes, a rich progress in theoretical and experimental studies was made over the past hundred years (Debye and Hückel,1923). The dielectric decrement of electrolyte solution with the salt concentration is an essential characteristic that governs the ion-ion, ion-solvent interactions.

Computable facts of the influence of ions on the dielectric constant of solution have been accepted as extremely important to the understanding of the forces operating in electrolyte solutions. Glueckauff (1964) developed a mathematical model treating the electrolytic solutions as a dielectric continuum and evaluated expressions for the lowering of dielectric constant of the electrolytic solution, on addition of the solute. Sastry (1970), Rao (1967) and Brahmajirao (1977) in their doctoral theses reported dielectric data of several electrolytic solutions. Other semi-empirical models such as the Pitzer (1961),Bromley (1973) and Davis (1993), and models related to other parameters have been developed to study the thermodynamic properties of electrolyte solution. Since 1961, Pitzerhas chosen the best available system proposed by Guggenheim(1935), developed in a series of papers , detailed mathematical equations for activity coefficient, osmotic coefficient and other thermodynamic parameters.

Still the permittivity parameter of the dielectric continuum is being left unaccounted in all the above attempts by several workers in the field. The first efficient investigational research of the dielectric properties of ionic solutions was presented by Hasted et al (1948). He observed that dielectric constant of a solution was reduced with the solute concentration. However, Glueckauff (1964) developed a model which is verified to be agreeing very well with the experimental data. He had given a theoretical expression for calculating the dielectric constant of the solvent at different concentration of the electrolyte. In addition to few earlier workers, Brahmajirao(1977) had experimentally determined the dielectric constants at different concentrations by means of the Resonance method (Bradley *et al.* 1979;Uematsu *et al.* 1980; Archer *et al.* 1990; Fernandez *et al.,* 1997)

Ball *et al*., (1985), Furst and Renon (1993) have developed the equations for dielectric constant for the concentrated solutions. In recent times, Levy, Andelman and Orland (2012) presented a model for the dielectric decrement in concentrated solutions. They evaluated the mean dielectric constant around each ion and accounted for ion-ion interactions. The dielectric constant is an essential parameter for the short range as well as long range interactions, since it governs the electrostatic behaviour. At higher concentrations substantial deviations from linearity are observed (Hasted *et al.* 1948).The comprehensive mathematical model incorporating the variation of dielectric constant of the electrolyte as a function of concentration of the solute in the solvent is still to be developed. In dilute solutions, the author proposed the dielectric constant decreases nonlinearly with concentration (Rajendra *et al.* 2013) as follows.

$$\varepsilon_r = \varepsilon_w - \lambda_1\sqrt{c} - \lambda_2 c - \lambda_3 c^{\frac{3}{2}}, \qquad (1)$$

where $\varepsilon_w$ is the dielectric constant of pure water, 'c' is the concentration of electrolyte in the solvent, and $\alpha$ is an excess polarization parameter of the ionic species.

Margaret (2007), and Robinson and Stokes (1959) pointed that the internal field around the central ion of the ionic atmosphere created by the charged solute ions in the solvent of the electrolytic solution, tend to reduce the value of the dielectric constant of the electrolytic solution. Several unsuccessful attempts were made so far to deduce a comprehensive model, and to modify the DH Theory (1923) for an electrolyte solution, and the need to develop a comprehensive mathematical model to incorporate the depression in the dielectric constant of the dielectric continuum, arising due to solute-solvent interactions (Brahmajirao *et al.* 1977).

**Electro-osmatic flow of aqueous electrolyte solution in nanochannel**

It is essential to study the electro-kinetic phenomena to develop a complete mathematical model of electrolyte solution in channels. The surface charges on the channel are required for the formation of the electrical double layer. Teorell – Sievers-Meyer model take up identical distribution of static charge, moveable ions, and electric potential (Hagmeyer *et al.* 1999, Bowen *et al.* 2002, Garcia-Aleman *et al.* 2004).

However, these conventions are not appropriate to electrolyte solutions with channel having a radius larger than the Debye length. This leads to the plan of the space-charge model by Gross and Osterle (1968), and its improvement by Wang *et al.* (1995). The validity of this model to describe the electro-kinetic phenomena in charged channels has been established (Szymczyk *et al.* 1999; Yaroshchuk *et al.* 2005).This model is calculation exhaustive, mathematically complex, and computationally expensive, which creates complexities that are hard to apply it to mixed electrolyte solution. This demanded to the development of easy methods for mathematical modeling of electrolyte solution through nano channel (Hawkins Cwirko and Carbonell, 1989; Basu *et al.* 1997).

The comprehensive mathematical model of electro-osmotic flow for an aqueous electrolyte solution can be developed by coupled system of Navier – Stokes – Maxwell – Stefan – Poisson - Boltzmann (NSMSPB) equations. The governing equations for electro-osmotic flow are consideration of three main aspects that require the modification. They are electric potential $(\Psi)$,mass transport $(X_i)$ and kinetic flow $(u)$. The Navier-Stokes equations represent the flow field. The Maxwell-Stefan equations model represents the diffusive transport of multi component ions and Poisson-Boltzmann equations describe the variation of electrical potential in an aqueous electrolyte solution.

The Electro-osmotic flow systems of equations are

$$\varepsilon_1^2 \frac{\partial^2 \Psi}{\partial X^2} + \varepsilon_2^2 \frac{\partial^2 \Psi}{\partial Y^2} + \frac{\partial^2 \Psi}{\partial Z^2} = -\frac{\eta}{\varepsilon^2} \sum_{i=1} Z_i X_i \quad (2)$$

$$\nabla^2 X_i + Z_i \nabla.(X_i \nabla \Psi) = 0 \quad (3)$$

$$\varepsilon_1^2 \frac{\partial^2 u}{\partial X^2} + \varepsilon_2^2 \frac{\partial^2 u}{\partial Y^2} + \frac{\partial^2 u}{\partial Z^2} = -\frac{\eta}{\varepsilon^2} \sum_{i=1} Z_i X_i \quad (4)$$

Subjected to the boundary conditions

$$Z = 0, \quad \Psi = \psi_0, \quad X_i = X_i^0, \quad u = u^1 \quad (5)$$

$$Z = 1, \quad \Psi = \zeta, \quad X_i = X_i^1, \quad u = u^2 \quad (6)$$

$$Y = 0, \quad \Psi = \psi_0, \quad X_i = X_i^0, \quad u = u^3 \quad (7)$$

$$Y = 1, \quad \Psi = \zeta, \quad X_i = X_i^0, \quad u = u^4, \quad (8)$$

where

$$\varepsilon = \frac{\lambda}{h} = \frac{\text{Debye length}}{\text{height of the nanochannel}} \text{ and } \eta = \frac{c}{I} = \frac{\text{Concentration of solute}}{\text{Ionic strength}}.$$

The variation of dielectric constant deduced above has to be incorporated at the appropriate places in the governing equations. The solutions of these equations which are satisfied at every point of the region R are subjected to certain boundary conditions. The problems in which all of the boundary conditions, relating to interior of a given region are of the essential type are called Dirichlet (Geometric) boundary value problems. The problems in which all of the boundary conditions are of the natural type are called Neumann (Dynamic) boundary value problems. Mixed boundary value problems are those in which both essential and natural boundary conditions are specified.

**One dimensional governing equations and finite difference method**

The electric potential, mass transport and flow field are very important physical quantities when studying electro-kinetics in an aqueous electrolyte solution. In our work, we used the rectangular nano-channel for which length and width are much larger compared to height ($L \gg h, w \gg h$). The author applied the fully coupled one dimensional Navier- Stokes- Maxwell-Stefan- Poisson-Boltzmann (NSMSPB) equations. The electrolytic system using finite difference method and the solutions were analyzed with the available dielectric data. Electrolytes like NaCl, KCl and $CaCl_2$ are binary electrolytes which on dissolution give one cation and one anion per molecule.

Consider the infinite parallel layers of nano-channel ($L \gg h, w \gg h$), so that the terms with coefficients $\varepsilon_1 = \frac{h}{L} \ll 1, \varepsilon_2 = \frac{h}{w} \ll 1$ become negligibly small in the governing equations. The equations (2), (3) and (4) become

$$\frac{d^2\Psi}{dZ^2} = -\frac{\eta}{\varepsilon^2}\sum_{i=1} Z_i X_i \tag{9}$$

$$\frac{d}{dZ}\left(\frac{\partial X_i}{\partial Z} + Z_i . X_i \frac{d\Psi}{dZ}\right) = 0 \tag{10}$$

$$\frac{d^2 u}{dZ^2} = -\frac{\eta}{\varepsilon^2}\sum_{i=1} Z_i X_i \tag{11}$$

Subjected to the boundary conditions

$$Z = 0, \qquad \Psi = \psi_0 \tag{12}$$

$$Z = 1, \qquad \Psi = \zeta \tag{13}$$

$$Z = 0, \qquad X_i = X_i^0 \tag{14}$$

$$Z = 1, \quad X_i = X_i^1 \tag{15}$$

$$Z = 0, \quad u = 0 \tag{16}$$

$$Z = 1, \quad u = 0 \tag{17}$$

The electro-chemical properties of parallel surfaces will be the same if the nano-channel is symmetric. Therefore, for symmetric nano-channels we take the following boundary conditions (assuming symmetric distribution of electrolytic ions).

$$Z = 0, \quad Z = 1, \quad \Psi = 0 \tag{18}$$

$$Z = 0, \quad Z = 1, \quad X_i = X_i^0 \tag{19}$$

$$Z = 0, \quad Z = 1, \quad u = 0 \tag{20}$$

The analytical methods of solving partial differential equations are applicable only to a limited class of equations. Quite often the partial differential equations appearing in physical problem as in our case do not belong to any of these familiar types and one has to resort to numerical methods. These methods are even of greater importance for the purpose of computation using the latest available software. The finite difference method is introduced in this section to solve the equations (9), (10) and (11) that represent a continuous mechanism with respect to space. To facilitate nearly exact evaluation with boundary conditions applied for specific elements considered in the system, the continuous equations get transformed into difference equations.

To obtain the appropriate finite difference approximations in regard to (i) electrical potential, (ii) charge transport and (iii) kinetic flow, the Taylor series expansion used with respect to the position 'x' for $\phi$ chosen to represent either of the above three variables is detailed below. The final equations of such treatment represent either of the above three variables.

Using Taylor series, the function $\phi\,(x + \Delta x)$ can be expanded as

$$\phi\,(x + \Delta x) = \phi(x) + \Delta x\,\phi'(x) + \frac{(\Delta x)^2}{2}\phi''(x) + \frac{(\Delta x)^3}{6}\phi'''(x) + \cdots \quad (21)$$

From which we obtain

$$\phi'(x) = \frac{\phi\,(x + \Delta x) - \phi(x)}{\Delta x}, \quad (22)$$

which is the forward difference approximation for $\phi'(x)$.
Similarly, by expansion of $\phi(x - \Delta x)$ in Taylor's series, we get

$$\phi\,(x - \Delta x) = \phi(x) - \Delta x\,\phi'(x) + \frac{(\Delta x)^2}{2}\phi''(x) - \frac{(\Delta x)^3}{6}\phi'''(x) + \cdots \quad (23)$$

Leading to

$$\phi'(x) = \frac{\phi(x) - \phi\,(x - \Delta x)}{\Delta x}, \quad (24)$$

which is the backward difference approximation for $\phi'(x)$.
A central difference approximation for $\phi'(x)$ can be obtained by subtractin (24) (22) from

i.e. $\phi'(x) = \frac{\phi\,(x + \Delta x) - \phi\,(x - \Delta x)}{2\,\Delta x}$ (25)

It is clear that (25) is better approximation to $\phi'(x)$ than either (22) or (24) because a central point projects correct evaluation of difference.
Similarly we get for $\phi''(x)$ by addition of (9) and (11),

$$\phi''(x) = \frac{\phi\,(x\,+\,\Delta x)\,-\,\phi(x)\,+\,\phi\,(x\,-\,\Delta x)}{(\Delta x)^2} \qquad (26)$$

To apply the above notation to the problem defined by (9), (10) and (11), we define

$$\phi\;=\;f\;(\,x, y, z\,\ldots\ldots)$$

in which variable 'x' varies from $x_0$ to $x_n$. Considering only one dimension for the evaluation of the differentials, other variables 'y', 'z'…treated as constants. Divide the range $[x_0, x_n]$ into 'n' equal subintervals of width $\Delta x$ so that

$$x_j = x_0\;+\;j\Delta x, \qquad j\;=\;0,1,2,3\ldots\ldots n, \qquad (27)$$

where $\Delta x\;\ll\;1$, is assumed to be very small elemental change in 'x'; 'j' is assumed to be an integer to count the number of such elements in the process of evaluation. This ensures the difference in evaluation to be uniform over the chosen element that is presumed to be very small. A diagrammatic representation of this is given in figure 1.

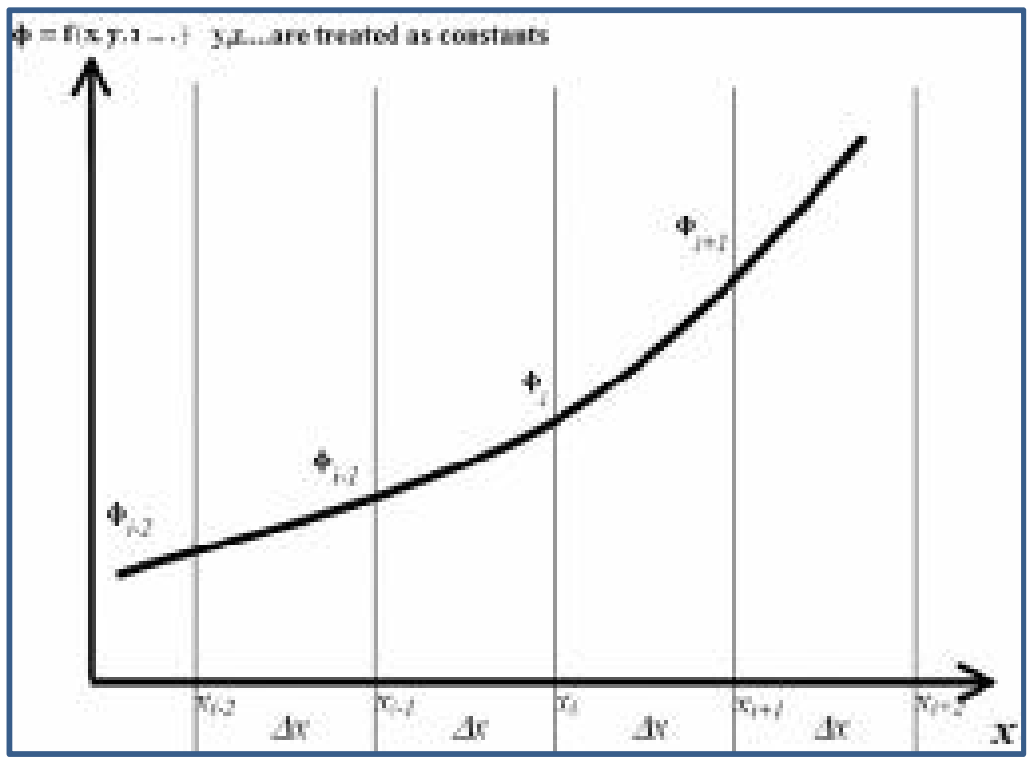

Figure 1. Function $\phi\;=\;f(x, y, z\,\ldots)$ in the sight of Finite difference method

The corresponding values of function $\phi$ at these points (elements)

$$\phi(x_j) = \phi_j = \phi(x_0 + j\Delta x),$$

$$j = 0,1,2,3, \dots\dots n \tag{28}$$

From the equations (25) and (26),

$$\left(\frac{d\phi}{dx}\right)_j = \frac{\phi_{j+1} - \phi_{j-1}}{2\Delta x} \quad \text{and} \tag{29}$$

$$\left(\frac{d^2\phi}{dx^2}\right)_j = \frac{\phi_{j+1} - 2\phi_j + \phi_{j-1}}{(\Delta x)^2} \tag{30}$$

Using these equations (29) and (30) with (9), (10) and (11) for $j = 0,1,2,3, \dots n$, we have

$$\Psi_{j+1} - 2\Psi_j + \Psi_{j-1} = -\frac{\eta(\Delta x)^2}{\varepsilon^2}\sum_{i=1} Z_i(X_i)_j \tag{31}$$

$$\frac{s(X_i)_{j+1} - 2\,(X_i)_j + (X_i)_{j-1}}{(\Delta x)^2} + Z_i\frac{(X_i)_{j+1} - (X_i)_{j-1}}{2\,\Delta x}\frac{\Psi_{j+1} - \Psi_{j-1}}{2\,\Delta x}$$

$$+ Z_i(X_i)_j\frac{\Psi_{j+1} - 2\Psi_j + \Psi\phi_{j-1}}{(\Delta x)^2} = 0 \tag{32}$$

$$u_{j+1} - 2u_j + u_{j-1} = -\frac{\eta(\Delta x)^2}{\varepsilon^2}\sum_{i=1} Z_i(X_i)_j, \tag{33}$$

where $X_i = \frac{c_i}{c}$, $(X_i)_j$ is mole fraction of ion species ‘i’ at $x = x_j$.

Using equation (31) with (32), we have

$$\frac{(X_i)_{j+1} - 2\,(X_i)_j + (X_i)_{j-1}}{(\Delta x)^2} + Z_i\,\frac{(X_i)_{j+1} - (X_i)_{j-1}}{2\,\Delta x}\frac{\Psi_{j+1} - \Psi_{j-1}}{2\,\Delta x}$$

$$+\mathrm{Z_i}(\mathrm{X_i})_j \frac{-\frac{\eta(\Delta x)^2}{\varepsilon^2}\sum_{\mathrm{i}=1}\mathrm{Z_i}(\mathrm{X_i})_j}{(\Delta x)^2} = 0$$

$$\Rightarrow (\mathrm{X_i})_{j+1} - 2\,(\mathrm{X_i})_j + (\mathrm{X_i})_{j-1} + \mathrm{Z_i}(\mathrm{X_i})_{j+1} - (\mathrm{X_i})_{j-1}\Psi_{j+1} - \Psi_{j-1}$$

$$= \mathrm{Z_i}(\mathrm{X_i})_j \frac{\eta(\Delta x)^2}{\varepsilon^2}\sum_{\mathrm{i}=1}\mathrm{Z_i}(\mathrm{X_i})_j = 0$$

$$\Rightarrow [1+\mathrm{Z_i}](\mathrm{X_i})_{j+1} - 2\,(\mathrm{X_i})_j + [1+\Psi_{j+1}-\Psi_{j-1}](\mathrm{X_i})_{j-1}$$

$$= \mathrm{Z_i}(\mathrm{X_i})_j \frac{\eta(\Delta x)^2}{\varepsilon^2}\sum_{\mathrm{i}=1}\mathrm{Z_i}(\mathrm{X_i})_j = 0 \quad (34)$$

Assuming the values of ‘Ψ’ and ‘u’ for the elements with i = 0 and i = n to be zero for initial and final elements, the value $(X_i)_0$ and $(X_i)_n$ reduce to the below form

$$\Psi_0 = 0, \qquad \Psi_n = 0 \quad (35)$$

$$(X_i)_0 = X_i^0, (X_i)_n = X_i^0 \quad (36)$$

$$u_0 = 0, \qquad u_n = 0 \quad (37)$$

Equation (31), (33) and (34) assume the form of

$$p_j\phi_{j-1} + q_j\phi_j + r_j\phi_{j+1} = d_j \quad (38)$$

Therefore, for j=1 to j=n, a system of linear equations for each of the three chosen variables are obtained.
Let us consider the system of linear equations for $\phi_j$

$$p_1\phi_0 + q_1\phi_1 + r_1\phi_2 = d_1 \quad (39)$$

$$p_2\phi_1 + q_2\phi_2 + r_2\phi_3 = d_2 \quad (40)$$

… … … … … … … … … … …

$$p_{j-1}\phi_{j-2} + q_{j-1}\phi_{j-1} + r_{j-1}\phi_j = d_{j-1} \quad (41)$$

$$p_j\phi_{j-1} + q_j\phi_j + r_j\phi_{j+1} = d_j \quad (42)$$

… … … … … … … … … … … … … … …

$$p_n\phi_{n-1} + q_n\phi_n + r_n\phi_{n+1} = d_n, \quad (43)$$

where$p_j,\ q_j,\ r_j\ ,\ d_j$ are the coefficients of system of linear equations. $\phi_j = \phi(x_j),\ j = 1,2,3\ldots.n$are values of $\phi$ which are unknown. The above system of equations is solved using Thomas-Algorithm, a technique based on Gauss-elimination. $\phi_0,\ \phi_n$are the initial and final values of $\phi$.
From the equation (39), we get

$$\phi_1 = \frac{\beta_1}{\alpha_1} - \frac{r_1\phi_2}{\alpha_1}, \quad (44)$$

where $\alpha_1 = q_1,\quad \beta_1 = d_1 - p_1\phi_0$. Substituting the equation (44) into the equation (40), we obtain

$$\alpha_2\phi_2 + r_2\phi_3 = \beta_2, \quad (45)$$

where$\alpha_2 = q_2 - \frac{p_2 r_1}{\alpha_2},\ \beta_2 = d_2 - \frac{p_2\,\beta_1}{\alpha_2}$.

From the equation (45), we get

$$\phi_2 = \frac{\beta_2}{\alpha_2} - \frac{r_2\phi_3}{\alpha_2} \quad (46)$$

By continuing this successive substitution process, we get

$$\alpha_j\phi_j + r_j\phi_{j+1} = \beta_j\ , \qquad j = 2,3,4\ldots n, \quad (47)$$

where $\alpha_j = q_j - \frac{p_j r_{j-1}}{\alpha_j}, \quad \beta_j = d_j - \frac{p_j \beta_{j-1}}{\alpha_j}$.

For $j = n - 1$, from the equation (43), we get

$$\phi_{n-1} = \frac{\beta_{n-1}}{\alpha_{n-1}} - \frac{r_{n-1}\phi_n}{\alpha_{n-1}} \quad (48)$$

$\phi_n$ is the boundary value which is known to us. Finally,

$$\phi_j = \frac{\beta_j}{\alpha_j} - \frac{r_j \phi_{j+1}}{\alpha_j}, \qquad j = n-1, n-2, \dots\dots .3,2,1 \quad (49)$$

Observed that our discretized equations (31), (33) and (34) for $\Psi$, $X_i$, $u$ form tri-diagonal system of linear equations. In order to solve them, the iteration process is done by setting initial assumed values based on the boundary conditions.

$$\Psi_j = \Psi_j^0, \qquad (X_i)_j = (X_i)_j^{\,0}, \qquad u_j = u_j^0,$$

$$j = 1,2,3 \dots\dots\dots n \quad (50)$$

Use $\Psi_j$, $(X_i)_j$ and $u_j$ values to determine the coefficients in discretized equations (31), (33) and (34) and Thomas-Algorithm can be applied to evaluate a fresh $\Psi_j$, $(X_i)_j$ and $u_j$ values.

$$\Psi_j = \Psi_j^1, \qquad (X_i)_j = (X_i)_j^{\,1}, \qquad u_j = u_j^1, \qquad j = 1,2,3 \dots\dots\dots n \quad (51)$$

Continue the iteration process till it converges. Suppose after ‘k’ iterations, we have

$$\left|\frac{\Psi_j^k - \Psi_j^{k-1}}{\Psi_j^k}\right| < \delta, \quad j = 1,2,3 \dots n \quad (52)$$

$$\left|\frac{{(X_i)_j}^k - {(X_i)_j}^{k-1}}{{(X_i)_j}^k}\right| < \delta, \quad i = 1,2, \ldots .N,$$

$$j = 1,2,3 \ldots n \tag{53}$$

$$\left|\frac{u_j^k - u_j^{k-1}}{u_j^k}\right| < \delta, \quad i = 1,2, \ldots .N,$$

$$j = 1,2,3 \ldots n, \tag{54}$$

where$\delta = 10^{-4}$, which is small number.
If the iteration process converges, then

$$\Psi_j = \Psi_j^k, (X_i)_j = {(X_i)_j}^k, \qquad u_j = u_j^k,$$

$$j = 1,2,3 \ldots n \tag{55}$$

is the solution (31), (33) and (34). Therefore$\Psi_j$, $(X_i)_j$ and $u_j$ are the numerical solutions of one dimensional governing equations.

**Two dimensional governing equations and finite difference method**

The rectangular nano-channel where $\varepsilon_1 = \frac{h}{L} \ll 0$ and $\varepsilon_2 = \frac{h}{w} \ll 0$ as demonstrated in figure 2, the electro-osmatic flow cannot be considered as one dimensional model. The two dimensional model governing equations obtained from (2), (3) and (4) are given below.

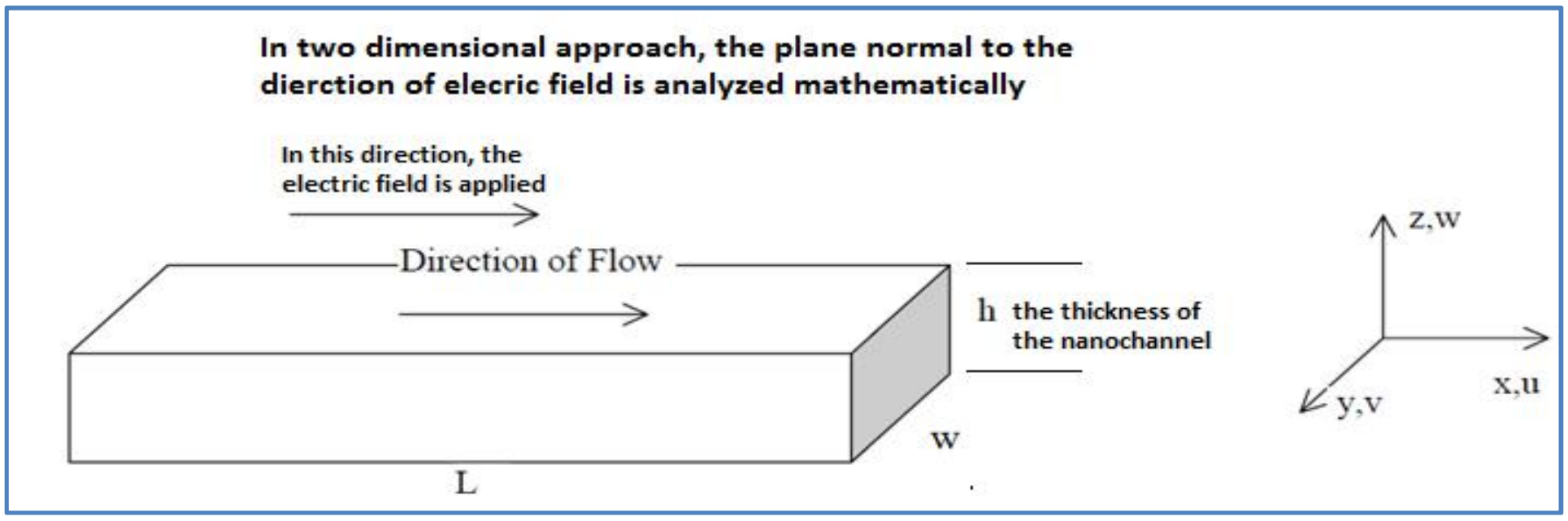

Figure 2. The rectangular nanochannel where $\varepsilon_1 = \frac{h}{L} \ll 0$

$$\varepsilon_2^2 \frac{\partial^2 \Psi}{\partial Y^2} + \frac{\partial^2 \Psi}{\partial Z^2} = -\frac{\eta}{\varepsilon^2} \sum_{i=1}^{N} Z_i X_i \tag{56}$$

$$\varepsilon_2^2 \frac{\partial}{\partial Y}\left(\frac{\partial X_i}{\partial Y} + Z_i . X_i \frac{\partial \Psi}{\partial Y}\right) + \frac{\partial}{\partial Z}\left(\frac{\partial X_i}{\partial Z} + Z_i . X_i \frac{\partial \Psi}{\partial Z}\right) = 0 \tag{57}$$

$$\varepsilon_2^2 \frac{\partial^2 u}{\partial Y^2} + \frac{\partial^2 u}{\partial Z^2} = -\frac{\eta}{\varepsilon^2} \sum_{i=1}^{N} Z_i X_i , \tag{58}$$

where $\varepsilon = \frac{\lambda}{h}$ , $\varepsilon_2 = \frac{h}{w}$ and $\eta = \frac{c}{I}$.

The boundary conditions of the equations (2), (3) and (4) become

$$Y = 0, \quad Z = 0, \quad \Psi = \psi_0 \tag{59}$$

$$Y = 1, \quad Z = 1, \quad \Psi = \zeta \tag{60}$$

$$Y = 0, \quad Z = 0, \quad X_i = X_i^0 \tag{61}$$

$$Y = 1, \quad Z = 1, \quad X_i = X_i^1 \tag{62}$$

$$Y = 0, \quad Z = 0, \quad u = 0 \tag{63}$$

$$Y = 1, \quad Z = 1, \quad u = 0 \tag{64}$$

The electro-chemical properties of parallel surfaces will be the same if the nano-channel is symmetric. Therefore, for symmetric nano-channels we take the following boundary conditions (assuming symmetric distribution of electrolytic ions).

$$Y = 0, \quad Y = 1, \quad Z = 0, \quad Z = 1, \quad \Psi = 0 \tag{65}$$

$$Y = 0, \quad Y = 1, \quad Z = 0, \quad Z = 1, \quad X_i = X_i^0 \tag{66}$$

$$\mathrm{Y} = 0,\ \mathrm{Y} = 1,\ \mathrm{Z} = 0,\ \mathrm{Z} = 1,\ u = 0 \tag{67}$$

The solutions of (56), (57) and (58) are satisfied at every point of the rectangular nano-channel subjected to Dirichlet (Geometric) boundary conditions. Using the Taylor series, the functions $\phi\,(x + \Delta x, y)$ and $\phi\,(x - \Delta x, y)$ can be expanded as

$$\phi\,(x + \Delta x, y) = \phi(x,y) + \Delta x \frac{\partial}{\partial x}\phi(x,y) + \frac{(\Delta x)^2}{2}\frac{\partial^2}{\partial x^2}\phi(x,y)$$

$$+ \frac{(\Delta x)^3}{6}\frac{\partial^3}{\partial x^3}\phi(x,y) + \cdots \tag{68}$$

$$\phi\,(x - \Delta x, y) = \phi(x,y) - \Delta x \frac{\partial}{\partial x}\phi(x,y) + \frac{(\Delta x)^2}{2}\frac{\partial^2}{\partial x^2}\phi(x,y)$$

$$- \frac{(\Delta x)^3}{6}\frac{\partial^3}{\partial x^3}\phi(x,y) + \cdots \tag{69}$$

A central difference approximation for $\frac{\partial \phi}{\partial x}$ can be obtained by subtracting (69) from(68) and neglect higher terms, i.e.,

$$\frac{\partial \phi}{\partial x} = \frac{\phi\,(x + \Delta x, y) - \phi\,(x - \Delta x, y)}{2\,\Delta x} + O(\Delta x^2) \tag{70}$$

Again, adding (68) and (69), we get the central difference approximation for $\frac{\partial^2 \phi}{\partial x^2}$, i.e.,

$$\frac{\partial^2 \phi}{\partial x^2} = \frac{\phi\,(x + \Delta x, y) - 2\,\phi(x,y) + \phi\,(x - \Delta x, y)}{(\Delta x)^2} + O(\Delta x^2) \tag{71}$$

Similarly, expand $\phi\,(x, y + \Delta y)$ and $\phi\,(x, y - \Delta y)$ by using Taylor's formula, we have

$$\phi\,(x, y + \Delta y) = \phi(x,y) + \Delta y \frac{\partial}{\partial y}\phi(x,y) + \frac{(\Delta y)^2}{2}\frac{\partial^2}{\partial y^2}\phi(x,y)$$

$$+ \frac{(\Delta y)^3}{6}\frac{\partial^3}{\partial y^3}\phi(x,y) + \cdots \tag{72}$$

Expand by the Taylor's formula, we have

$$\phi\,(x, y-\Delta y) = \phi(x,y) - \Delta y\,\frac{\partial}{\partial y}\phi(x,y) + \frac{(\Delta y)^2}{2}\frac{\partial^2}{\partial y^2}\phi(x,y)$$

$$-\frac{(\Delta y)^3}{6}\frac{\partial^3}{\partial y^3}\phi(x,y) + \cdots \tag{73}$$

A central difference approximation for $\frac{\partial \phi}{\partial y}$ can be obtained by subtracting (73) from(72) and neglect higher terms, i.e.,

$$\frac{\partial \phi}{\partial y} = \frac{\phi\,(x, y+\Delta y) - \phi\,(x, y-\Delta y)}{2\,\Delta y} + O(\Delta y^2) \tag{74}$$

Again, adding (72) and (73), we get an approximation for $\frac{\partial^2 \phi}{\partial y^2}$

$$\frac{\partial^2 \phi}{\partial y^2} = \frac{\phi\,(x, y+\Delta y) - 2\,\phi(x,y) + \phi\,(x, y-\Delta y)}{(\Delta y)^2} + O(\Delta y^2) \tag{75}$$

The values of $\frac{\partial \phi}{\partial x}$ and $\frac{\partial^2 \phi}{\partial x^2}$ at the point $\phi(x_i, y_j)$ can be now written as

$$\left(\frac{\partial \phi}{\partial x}\right)_{i,j} = \frac{\phi_{i+1,j} - \phi_{i-1,j}}{2\,\Delta x} + O(\Delta x^2) \tag{76}$$

$$\left(\frac{\partial^2 \phi}{\partial x^2}\right)_{i,j} = \frac{\phi_{i+1,j} - 2\phi_{i,j} + \phi_{i-1,j}}{(\Delta x)^2} + O(\Delta x^2) \tag{77}$$

Similarly, the values of $\frac{\partial \phi}{\partial y}$ and $\frac{\partial^2 \phi}{\partial y^2}$ at the point $\phi(x_i, y_j)$ can be written as

$$_{i,j} = \frac{\phi_{i,j+1} - \phi_{i,j-1}}{2\,\Delta y} + O(\Delta y^2) \quad (78)$$

$$\left(\frac{\partial^2 \phi}{\partial y^2}\right)_{i,j} = \frac{\phi_{i,j+1} - 2\phi_{i,j} + \phi_{i,j-1}}{(\Delta y)^2} + O(\Delta y^2) \quad (79)$$

The significance of the above mathematical equations can be diagrammatically explained using the rectangular cross section of the nanochannel shown below. The co-ordinates of the shown plotted points indicate the notation used in the development of the above equations. Finite difference approximations to derivatives $\frac{\partial \phi}{\partial x}$ , $\frac{\partial^2 \phi}{\partial x^2}$, $\frac{\partial \phi}{\partial y}$ and $\frac{\partial^2 \phi}{\partial y^2}$ are depicted in figure 3.

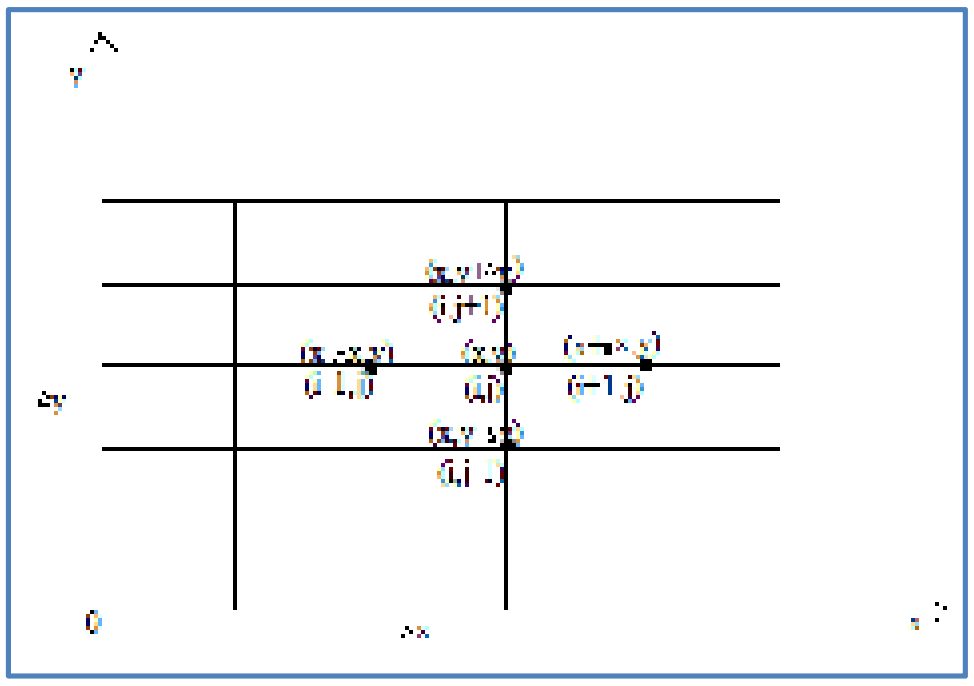

Figure 3. Finite difference approximations to derivatives $\frac{\partial \phi}{\partial x}$ , $\frac{\partial^2 \phi}{\partial x^2}$, $\frac{\partial \phi}{\partial y}$ and $\frac{\partial^2 \phi}{\partial y^2}$

We now obtain the finite-difference analogues of partial differential equations by replacing the derivatives in any equation by their corresponding difference approximations given above. Thus the Laplace equation in two dimensions, is

$$\frac{\partial^2 \phi}{\partial x^2} + \frac{\partial^2 \phi}{\partial y^2} = 0 \quad (80)$$

It has finite difference analogue

$$\frac{\phi_{i+1,j} - 2\phi_{i,j} + \phi_{i-1,j}}{(\Delta x)^2} + \frac{\phi_{i,j+1} - 2\phi_{i,j} + \phi_{i,j-1}}{(\Delta y)^2} = 0 \quad (81)$$

If $\Delta x = \Delta y$, we obtain

$$\phi_{i,j} = \frac{1}{4}\left(\phi_{i+1,j} + \phi_{i-1,j} + \phi_{i,j+1} + \phi_{i,j-1}\right) \tag{82}$$

The equation (82) shows that the value of $\phi$ at any point is the mean of its values at the four neighboring points. Let $\phi_{i,j}^{n}$ denotes the n$^{th}$ iterative value of $\phi_{i,j}$. An iteration procedure to solve (82) is

$$\phi_{i,j}^{(n+1)} = \frac{1}{4}\left(\phi_{i-1,j}^{(n)} + \phi_{i+1,j}^{(n)} + \phi_{i,j-1}^{(n)} + \phi_{i,j+1}^{(n)}\right) \tag{83}$$

for the interior mesh points and this is called Jacobi method. The method uses the latest iterative values available and scan the mesh points systematically from left to right along successive rows is known as Gauss Seidel method. The iterative formula for Gauss Seidel method is:

$$\phi_{i,j}^{(n+1)} = \frac{1}{4}\left(\phi_{i-1,j}^{(n+1)} + \phi_{i+1,j}^{(n)} + \phi_{i,j-1}^{(n+1)} + \phi_{i,j+1}^{(n)}\right) \tag{84}$$

It can be observed that the Gauss Seidel scheme converges twice as fast as the Jacobi scheme. Therefore, the equations (56), (57) and (58) are solved by applying Gauss Seidel method.

The equations (56), (57) and (58) are for the functions Ψ, $X_i$, $u$ and its variables Y, Z are dimensionless. (M+1), (N+1) mesh points on Y and Z axis respectively are chosen for discretizing the equations. This region is divided into a network of square mesh of side $\Delta Y = \Delta Z$ (assume that an exact subdivision of rectangle is possible).
Replacing the derivatives in (56) by their difference approximation, we obtain

$$\varepsilon_2^2\left(\frac{\Psi_{j,k+1} - 2\,\Psi_{j,k} + \Psi_{j,k-1}}{(\Delta Y)^2}\right) + \frac{\Psi_{j,k+1} - 2\,\Psi_{j,k} + \Psi_{j,k-1}}{(\Delta Z)^2}$$

$$= -\frac{\eta}{\varepsilon^2}\sum_{i=1}^{N} Z_i(X_i)_{j,k} \tag{85}$$

$$\Rightarrow \varepsilon_2^2\left(\frac{\Psi_{j,k+1} - 2\,\Psi_{j,k} + \Psi_{j,k-1}}{(\Delta Z)^2}\right) + \frac{\Psi_{j,k+1} - 2\,\Psi_{j,k} + \Psi_{j,k-1}}{(\Delta Z)^2}$$

$$= \ -\frac{\eta}{\varepsilon^2}\sum_{\mathrm{i}=1}^{\mathrm{N}} \mathrm{Z_i}(\mathrm{X_i})_{j,k} \qquad (86)$$

$$\Rightarrow \varepsilon_2^2\left(\Psi_{j,k+1} - 2\,\Psi_{j,k} + \Psi_{j,k-1}\right) + \ \Psi_{j,k+1} - + \Psi_{j,k-1}$$

$$= \ -\ \frac{\eta(\Delta Z)^2}{\varepsilon^2}\sum_{\mathrm{i}=1}^{\mathrm{N}} \mathrm{Z_i}(\mathrm{X_i})_{j,k} \qquad (87)$$

$$\therefore\ \Psi_{j,k}$$

$$= \frac{1}{4}\left(\varepsilon_2^2\Psi_{j,k+1} + \ \varepsilon_2^2\,\Psi_{j,k-1} + \ \Psi_{j,k+1} + \ \Psi_{j,k-1}\right.$$

$$\left. + \ \frac{\eta(\Delta Z)^2}{\varepsilon^2}\sum_{\mathrm{i}=1}^{\mathrm{N}} \mathrm{Z_i}(\mathrm{X_i})_{j,k}\right) \qquad (88)$$

Replacing the derivatives in (57) by their difference approximation, we obtain

$$\varepsilon_2^2\left(\frac{(\mathrm{X_i})_{j+1,k} - 2\,(\mathrm{X_i})_{j,k} \ + \ (\mathrm{X_i})_{j-1,k}}{(\Delta Y)^2}\right)$$

$$+ \ \varepsilon_2^2\mathrm{Z_i}\ \frac{(\mathrm{X_i})_{j+1,k} - (\mathrm{X_i})_{j-1,k}}{2\,\Delta Y}\frac{\Psi_{j+1,k} - \Psi_{j-1,k}}{2\,\Delta Y} + \frac{(\mathrm{X_i})_{j,k+1} - 2\,(\mathrm{X_i})_{j,k} \ + (\mathrm{X_i})_{j,k-1}}{(\Delta Z)^2}$$

$$+ \ \mathrm{Z_i}\ \frac{(\mathrm{X_i})_{j,k+1}(\mathrm{X_i})_{j,k-1}}{2\,\Delta Z}\frac{\Psi_{j,k+1} - \Psi_{j,k-1}}{2\,\Delta Z}$$

$$= \ 0 \qquad (89)$$

$$\Rightarrow \varepsilon_2^2\left(\frac{(X_i)_{j+1,k} - 2\,(X_i)_{j,k} + (X_i)_{j-1,k}}{(\Delta Z)^2}\right) + \varepsilon_2^2 Z_i\,\frac{(X_i)_{j+1,k} - (X_i)_{j-1,k}}{2\,\Delta Z}\,\frac{\Psi_{j+1,k} - \Psi_{j-1,k}}{2\,\Delta Z}$$

$$+\frac{(X_i)_{j,k+1} - 2\,(X_i)_{j,k} + (X_i)_{j,k-1}}{(\Delta Z)^2}$$

$$+Z_i\,\frac{(X_i)_{j,k+1}(X_i)_{j,k-1}}{2\,\Delta Z}\,\frac{\Psi_{j,k+1} - \Psi_{j,k-1}}{2\,\Delta Z}$$

$$= 0 \qquad (90)$$

$$\Rightarrow \varepsilon_2^2\big((X_i)_{j+1,k} - 2\,(X_i)_{j,k} + (X_i)_{j-1,k}\big)$$

$$+\,\varepsilon_2^2\Big(\big(Z_i\big((X_i)_{j+1,k} - (X_i)_{j-1,k}\big)\big)\big(\Psi_{j+1,k} - \Psi_{j-1,k}\big)\Big) + (X_i)_{j,k+1} - 2(X_i)_{j,k}$$

$$+\,(X_i)_{j,k-1}$$

$$+Z_i\big((X_i)_{j,k+1} - (X_i)_{j,k+1}\big)\big(\Psi_{j,k+1} - \Psi_{j,k-1}\big) = 0 \qquad (91)$$

$$\therefore (X_i)_{j,k} = \frac{1}{4}\Big(\varepsilon_2^2\big((X_i)_{j+1,k} + (X_i)_{j-1,k}\big)$$

$$+\,\varepsilon_2^2\Big(\big(Z_i\big((X_i)_{j+1,k} - (X_i)_{j-1,k}\big)\big)\big(\Psi_{j+1,k} - \Psi_{j-1,k}\big)\Big) + (X_i)_{j,k+1}$$

$$+\,(X_i)_{j,k-1}$$

$$+\,Z_i\big((X_i)_{j,k-1}(X_i)_{j,k-1}\big)\big(\Psi_{j,k+1}$$

$$-\,\Psi_{j,k-1})\Big) \qquad (92)$$

Replacing the derivatives in (58) by their difference approximation, we obtain

$$\varepsilon_2^2\left(\frac{u_{j,k+1}-2\,u_{j,k}+u_{j,k-1}}{(\Delta Y)^2}\right)+\frac{u_{j,k+1}-2\,\mathrm{u}_{j,k}+u_{j,k-1}}{(\Delta Z)^2}$$

$$=-\frac{\eta}{\varepsilon^2}\sum_{\mathrm{i}=1}^{\mathrm{N}}\mathrm{Z_i}(\mathrm{X_i})_{j,k} \tag{93}$$

$$\Rightarrow \varepsilon_2^2\left(\frac{u_{j,k+1}-2\,\mathrm{u}_{j,k}+u_{j,k-1}}{(\Delta Z)^2}\right)+\frac{u_{j,k+1}-2\,\mathrm{u}_{j,k}+u_{j,k-1}}{(\Delta Z)^2}$$

$$=-\frac{\eta}{\varepsilon^2}\sum_{\mathrm{i}=1}^{\mathrm{N}}\mathrm{Z_i}(\mathrm{X_i})_{j,k} \tag{94}$$

$$\Rightarrow \varepsilon_2^2\left(u_{j,k+1}-2\,\mathrm{u}_{j,k}+\mathrm{u}_{j,k-1}\right)+\mathrm{u}_{j,k+1}-2\,\mathrm{u}_{j,k}$$

$$+\,u_{j,k-1}$$

$$=-\frac{\eta(\Delta Z)^2}{\varepsilon^2}\sum_{\mathrm{i}=1}^{\mathrm{N}}\mathrm{Z_i}(\mathrm{X_i})_{j,k} \tag{95}$$

$$\therefore\ u_{j,k}=\frac{1}{4}\Bigg(\varepsilon_2^2 u_{j,k+1}+\ \varepsilon_2^2\,\mathrm{u}_{j,k-1}+\ u_{j,k+1}+\ \mathrm{u}_{j,k-1}$$

$$+\ \frac{\eta(\Delta Z)^2}{\varepsilon^2}\sum_{\mathrm{i}=1}^{\mathrm{N}}\mathrm{Z_i}(\mathrm{X_i})_{j,k}\Bigg) \tag{96}$$

The equations (88), (92) and (96) are solved by an iteration process by setting initial assumed values based on the boundary conditions.

$$\Psi_{j,k} = \Psi_{j,k}^{0}, \qquad (\mathrm{X_i})_{j,k} = \mathrm{X_i}_{j,k}^{0}, \qquad u_{j,k} = u_{j,k}^{0},$$

$$j = 1,2,3 \ldots\ldots\ldots M, \quad k = 1,2,3 \ldots\ldots\ldots N \tag{97}$$

The values of $\Psi_{j,k}$, $(\mathrm{X_i})_{j,k}$, and $u_{j,k}$ determine the coefficients in discretized equations (88), (92) and (96). Gauss-Seidel method is applied to evaluate a fresh $\Psi_{j,k}$, $(\mathrm{X_i})_{j,k}$, and $u_{j,k}$values.

$$\Psi_{j,k} = \Psi_{j,k}^{1}, \qquad (\mathrm{X_i})_{j,k} = \mathrm{X_i}_{j,k}^{1}, \qquad u_{j,k} = u_{j,k}^{1},$$

$$j = 1,2,3 \ldots\ldots\ldots M, \quad k = 1,2,3 \ldots\ldots\ldots N \tag{97a}$$

This iteration process is continued till the final result converges with the specified least possible error '$\delta$'. Suppose after '$l$' iterations, we have

$$\left|\frac{\Psi_{j,k}^{l} - \Psi_{j,k}^{l-1}}{\Psi_{j,k}^{l}}\right| < \delta, \quad J = 1,2, \ldots . M, k = 1,2,3 \ldots\ldots\ldots N, l = 0,1,2,3 \ldots n \tag{98}$$

$$\left|\frac{(\mathrm{X_i})_{j,k}^{l} - (\mathrm{X_i})_{j,k}^{l-1}}{(\mathrm{X_i})_{j,k}^{l}}\right| < \delta, \quad i = 1,2, \ldots . \mathrm{N}, \qquad J = 1,2, \ldots . M,$$

$$k = 1,2,3 \ldots\ldots\ldots N, \quad l = 0,1,2,3 \ldots n \tag{99}$$

$$\left|\frac{u_{j,k}^{l} - u_{j,k}^{l-1}}{u_{j,k}^{l}}\right| < \delta, \quad J = 1,2, \ldots .M, k = 1,2,3 \ldots \ldots \ldots N, l$$

$$= 0,1,2,3 \ldots n, \qquad (100)$$

where $\delta = 10^{-4}$ which is small number, and$\Delta Y = \Delta Z$ is used in the calculations. If the iteration process converges, then

$$\Psi_{j,k} = \Psi_{j,k}^{l}, (\mathrm{X_i})_{j,k} = (\mathrm{X_i})_{j,k}^{l}, \qquad u_{j,k} = u_{j,k}^{l},$$

$$j = 1,2,3 \ldots n \qquad (101)$$

is the solution of (88), (92) and (96). Therefore $\Psi_{j,k}$, $(\mathrm{X_i})_{j,k}$ and $u_{j,k}$ are the numerical solutions of two dimensional governing equations.

**Discussion**

**One Dimensional:** If the surface mole fractions are considered as constants for $Na^+$, $Cl^-$ions and $\varepsilon \ll 1$, then, the plots in figures 4 and 5 show the electrical potential and ionic velocity distribution of EOF across the nano-channel. The electrical potential '$\Psi$' and velocity '$u$'can be represented by similar curves. Clearly along the axis of the nanochannel, the electrical potential '$\Psi$' and velocity '$u$' have maximum value because of the motion of the charged particles is under consideration in the system. In presence of the electric field the bulk liquid occupies the region around the axis at which the solution is neutral because of neutralized charge content at that point. As we move from the axis towards the layer constituting the hypothetical wall of the nanochannel, we can find that the charge content gradually enhances due to ionization. Therefore it is expected for the '$\Psi$' and '$u$'values to be maximum about the axis. As the value of (h = 20 nm and h = 4 nm) height increases, we find that the potential distribution modifies from the shape of parabola to record nearly constant potential about the axis and falls to the least value about the wall of the nanochannel. For lesser value of 'h' the variation of $\Psi$' and '$u$' from axis to the wall becomes conspicuous in the continuous change. The valuesof $\Psi$' and '$u$' become unique at the axis. These observations have an intimated correlation with the dielectric permittivity and molarity of charge at the surfaces of the nanochannel under the consideration.

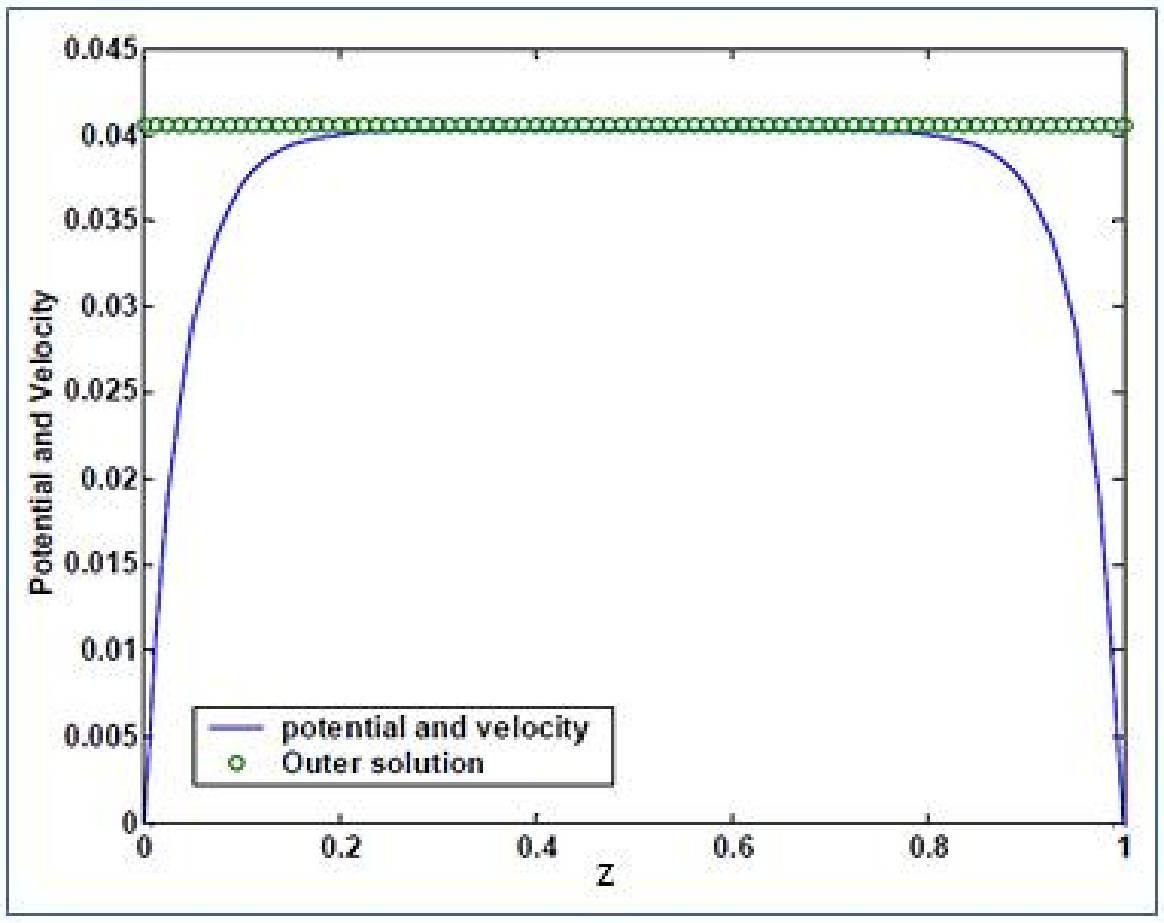

Figure 4. The electric potential 'Ψ'and velocity '$u$' distribution for NaCl in nano-channel. ($Na^+$ = 0.00276, $Cl^-$ = 0.00254, h=20 nm, $\varepsilon = \frac{\lambda}{h} = 0.04$)

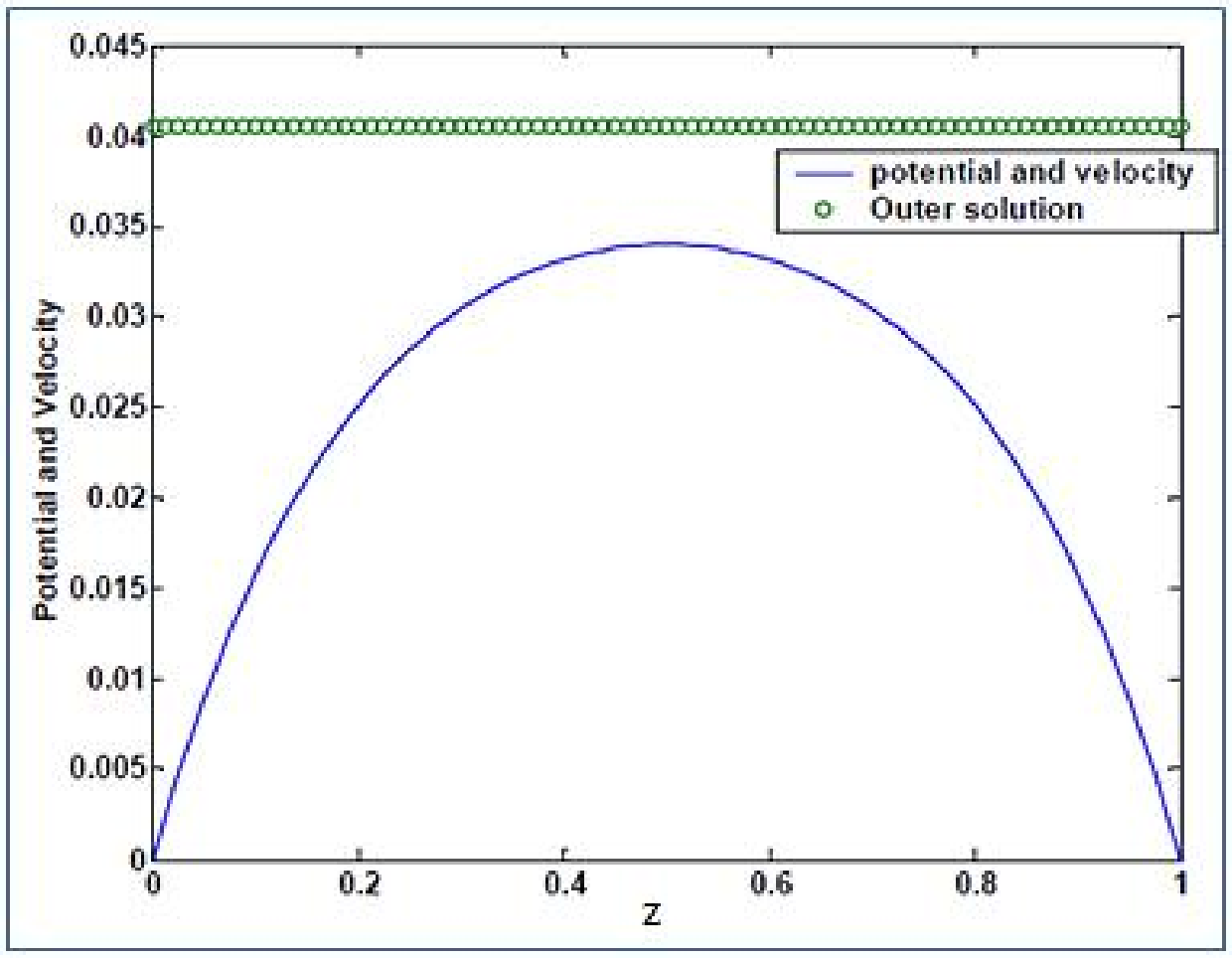

Figure 5. The electric potential 'Ψ'and velocity '$u$' distribution for NaCl in nano-channel. ($Na^+$ = 0.00276, $Cl^-$ = 0.00254, h=4 nm, $\varepsilon = \frac{\lambda}{h} = 0.2$)

According to the basic model of Debye-Hückel ,

$$\lambda = \frac{1}{\kappa} = \sqrt{\frac{\varepsilon_0\, \varepsilon_r k_B T}{e^2 \sum_{i=1}^{N} z_i^2 n_i'}} = \sqrt{\frac{\varepsilon_0\, \varepsilon_r\, R\, T}{F^2 \sum_{i=1}^{N} z_i^2 c_i}} = \frac{1}{F}\sqrt{\frac{\varepsilon_0 \varepsilon_r\, R\, T}{2\, I}}.$$

This equation expects reduction in λ as $\varepsilon_r$ suffers decrement with concentration of electrolyte. For a given value of 'h', consequently $\varepsilon$ decreases as $\varepsilon_r$ decreases. Thus, at a given value of 'h', $\varepsilon_r$ decreases '$\varepsilon$', tending to produce more variation in the potential and velocity. This fact is exactly reproduced in the plots. Thus the reduction in dielectric constant is solely responsible for a given nanochannel to reorganize the potential and velocity distribution about the axis of the channel. As the value of 'h' increases from 4 nm to 20 nm,the value of $\varepsilon$ decreases. Correspondingly for h = 20nm, both 'Ψ' and 'u' record very fast reduction in their value with respect to the axis of the channel. Hence, the potential and velocity distribution record a continuous change from the value at the axis resulting in parabolic variation as seen in the graph.

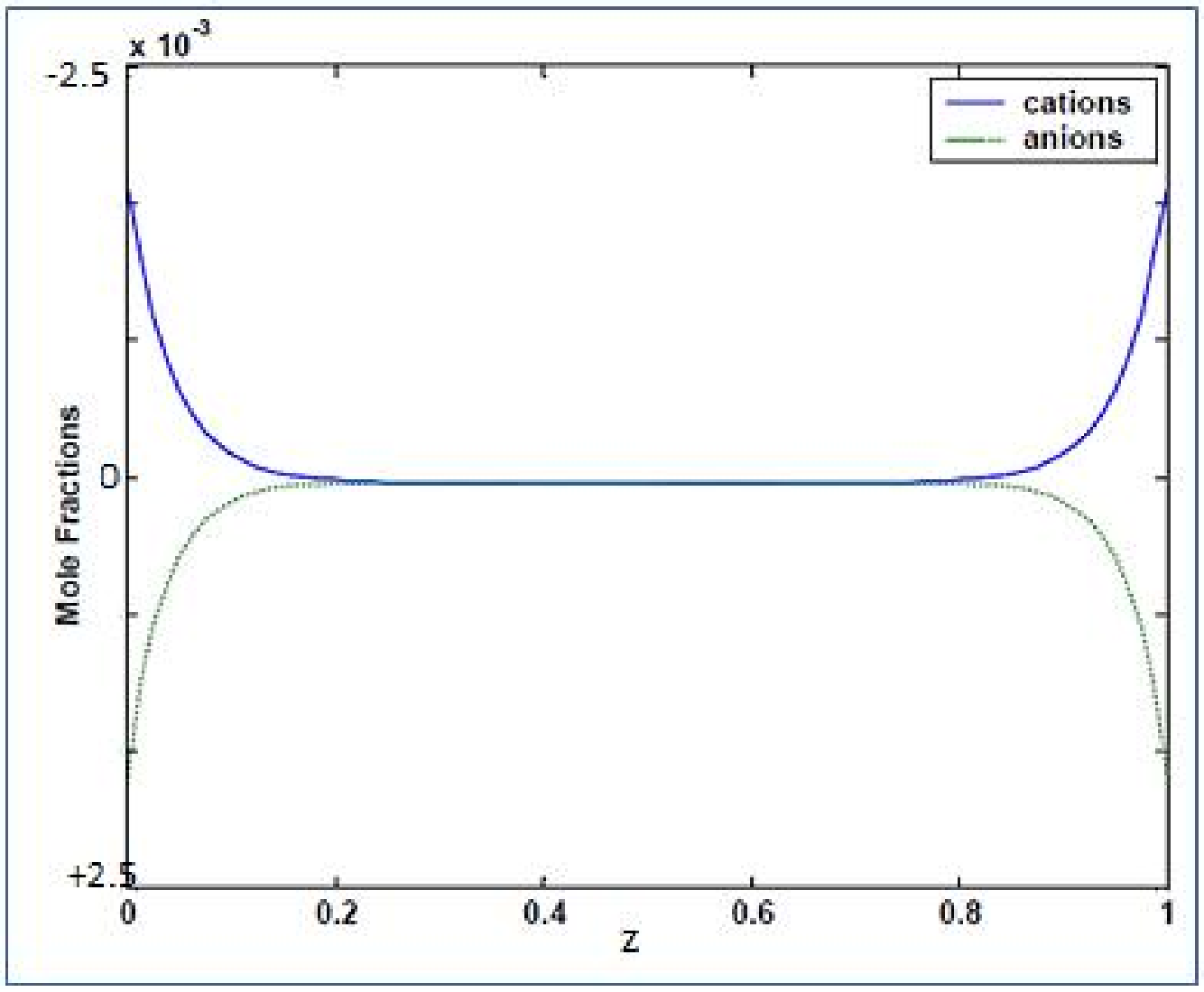

Figure 6. The mole fraction plot in nano-channel.
($Na^+$ = 0.00276, $Cl^-$ = 0.00254, h=20 nm, $\varepsilon = \frac{\lambda}{h} = 0.04$)

Figure 6 is a demonstration of the variation of mole fraction of distribution of $Na^+$, $Cl^-$ electrolyte ions in nano-channel along the Z-axis. Along the Z-axis and both $Na^{+}$, $Cl^{-}$ mole fractions $(X_{+} \& X_{-})$ are dimensionless quantities. In the above partial differential equations, the partial derivatives in the direction of Z-axis alone are considered because in this chapter the author attempted to solve the equations of the developed model for one dimension only. In the bulk solution, both mole fractions are equal because the aqueous electrolyte solution is neutral on the whole. For the electrical double layer made up of the anion and cation structures, '$\kappa$' represents the reciprocal of thickness of ionic atmosphere, hypothesized by Debye Hückel in their basic theory. '$\varepsilon$' the fundamental parameter governs the space available for the motion of charges through the channel. It was shown that '$\varepsilon$' is governed by simultaneous variations of '$\lambda$' and 'h'. Thus, the variation of molarity in the 'Z' direction represented by $\frac{\partial X_{+}}{\partial Z}$ and $\frac{\partial X_{-}}{\partial Z}$ is directly proportional to $\frac{\partial \Psi}{\partial Z}$. Hence, if the distance to the surface of the wall decreases, then the mole fraction of $Na^{+}$ and $Cl^{-}$ increases. However, the location of the layers is in opposite direction with respect to Z-axis. This expectation is in exact agreement with the plots (Fig.6 and 7).

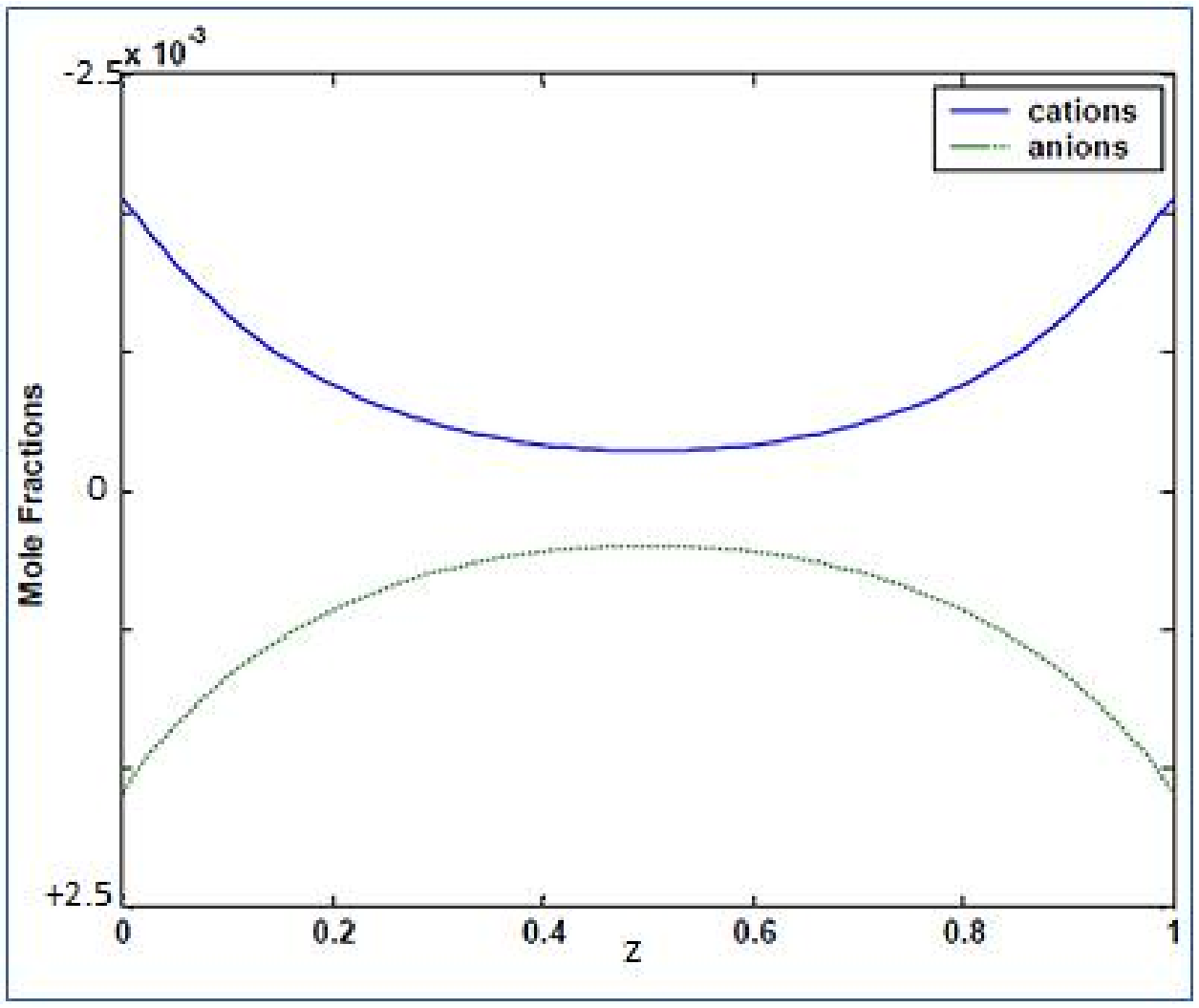

Figure 7. The mole fraction plot in nano-channel. Wall mole fractions are constants ($Na^{+}$ = 0.00276, $Cl^{-}$ = 0.00254, h=4 nm, $\varepsilon = \frac{\lambda}{h} = 0.2$)

Consider the situation where $\varepsilon = O(1)$, i.e., when 'h' is approximately equals to '$\lambda$', the channel thickness becomes approximately equal to the Debye length. At this critical condition the variation of molarity of ionic concentration assumes a saturation value defined by other physiochemical parameters of the electrolyte solution. Hence, it is found in the cited plot (Fig 7), the nano-channel restricts any possible change in the molarity of the ion corresponding to the value of ‘h’.

**Two dimensional:** In two dimensional nano-channel, the author assumes that $\frac{h}{L} \ll 0$. Therefore, the author chose three nanochannels with different geometry;

$$(i) 3.6\,\mu m\ X\ 20\,nm\ X\ 20\,nm$$
$$(ii) 3.6\,\mu m\ X\ 4\,nm\ X\ 4\,nm$$
$$(iii)\ 3.6\,\mu m\ X\ 20\,nm\ X\ 4\,nm,$$

that satisfies the condition $\frac{h}{L} \ll 0$. The author studied the chosen electrolytes in aqueous solution. The Debye length $\lambda = 0.79\,nm$ is same for all the above cases. We calculated $\varepsilon = \frac{\lambda}{h}$ value for different geometry nano-channels and shown the results in table 1.

Table 1. Three different dimensions of nano-channels having equal length, $L = 3.6\,\mu m$.

| | Height (nm) | Width (nm) | Debye length($\boldsymbol{\lambda}$) | $\boldsymbol{\varepsilon = \frac{\lambda}{h}}$ |
|---|---|---|---|---|
| Nanochannel 1 | 20 | 20 | 0.79 | 0.040 |
| Nanochannel 2 | 4 | 4 | 0.79 | 0.020 |
| Nanochannel 3 | 4 | 20 | 0.79 | 0.020 |

The results of electrical potential, velocity and mole fractions are checked by matching two mesh results (Table 2). The two meshes chosen are $41\,X\,41$ and $81\,X\,81$ in terms mesh points considered. The coordinates of the points in evaluation (3,3), (5,5) and (21,21) in $41\,X\,41$ mesh and (5,5), (9,9) and (41,41) in $81\,X\,81$ mesh clearly indicate very close agreement for the values of the parameters $X_1$, $X_2$, $\Psi$ and $u$ ($X_1$, $X_2$ indicate cation and anion values). These results on observation reveal that for co-ordinates (Y, Z) for $41\,X\,41$mesh nearly show total agreement with the co-ordinates$(2Y - 1,\ 2Z - 1)$ for $81\,X\,81$ mesh.

Table 2. The solution of $X_i$ , $\Psi$ and $u$at three points in $41\ X\ 41$, $81\ X\ 81$ mesh points

| 41×41 mesh points | | 81×81 mesh points | |
|---|---|---|---|
| (3.3) | $X_1$=0.5089<br>$X_2$=0.4904<br>Ψ=0.0220<br>u=0.0220 | (5.5) | $X_1$=0.5088<br>$X_2$=0.4906<br>Ψ=0.0224<br>u=0.0224 |
| (5.5) | $X_1$=0.5026<br>$X_2$=0.4966<br>Ψ=0.0345<br>u=0.0345 | (9.9) | $X_1$=0.5025<br>$X_2$=0.4967<br>Ψ=0.0347<br>u=0.0347 |
| (21.21) | $X_1$=0.4996<br>$X_2$=0.4996<br>Ψ=0.0406<br>u=0.0406 | (41.41) | $X_1$=0.4996<br>$X_2$=0.4996<br>Ψ=0.0408<br>u=0.0408 |

The errors between $41\ X\ 41$, $81\ X\ 81$ mesh points for the nanochannel 1 is less than 0.0002 for all electrical potential, velocity and mole fractions variables. This leads to the conclusion that if the number of mesh points chosen for evaluation become reasonably large, then the error for the values of the variables becomes insignificant. Also the variation observed in the values becomes unnoticeable when $41\ X\ 41$ mesh is changed to $81\ X\ 81$ mesh. Hence we can conclude that the values of $X_1$ , $X_2$, $\Psi$ and $u$ of $41\ X\ 41$ mesh can be taken to be precise.

The dimensionless electrical potential and velocity variations in the nanochannel 1 are displayed in figure8. The results show that electrical potential and velocity overlap each other in the graph, since 'Ψ' and 'u' have been deduced starting with identical governing equations and boundary conditions. Since the nanochannel 1 is designated as a square nanochannel, the Ψ' and 'u' variations in Y and Z direction are equal as shown in figure8.

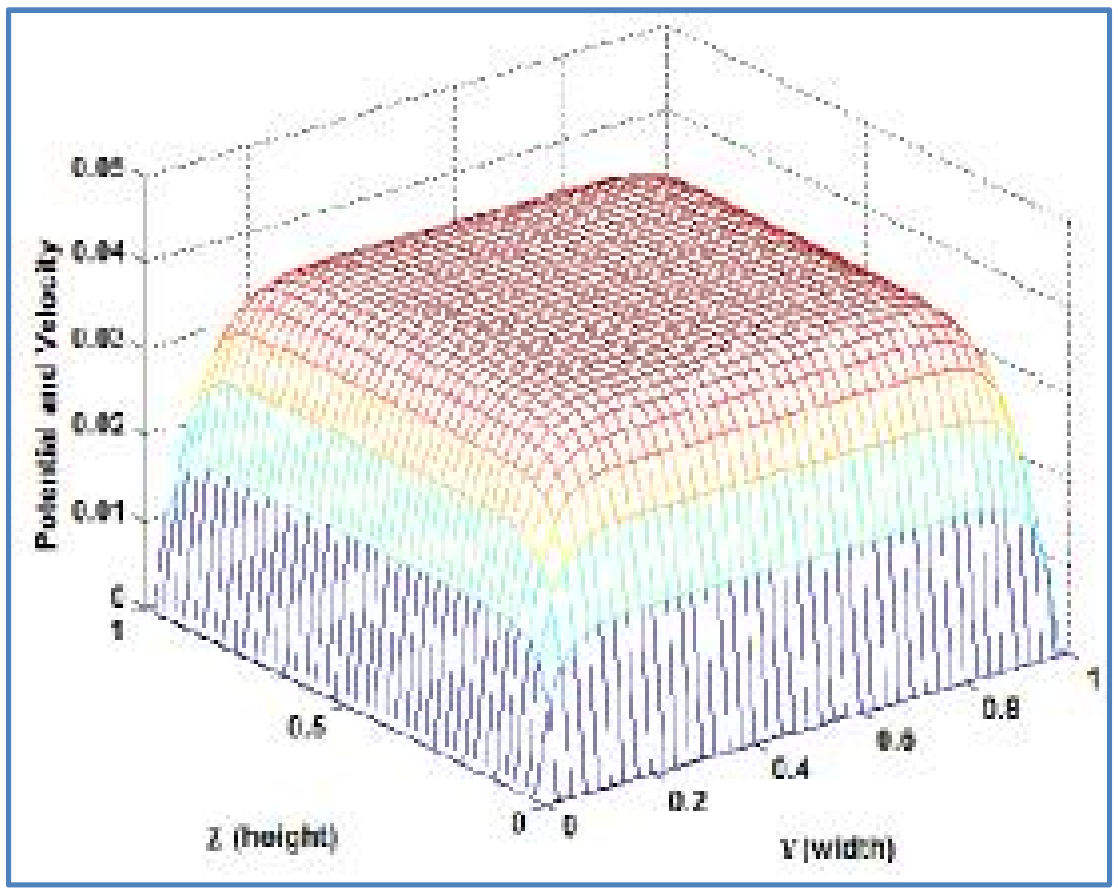

Figure 8. Dimensionless 'Ψ' and 'u' distribution for an aqueous electrolyte solution in nanochannel 1 with height 20 nm and width 20 nm.

The mole fractions of cation and anion species in nanochannel 1 are displayed in figure9. In the middle of the nanochannel, the mole fractions of both the species are same. Therefore, the solution is electrically neutral in the centre of the channel. In the region near to the boundary of the channel, the deposition of cations and anions take place.

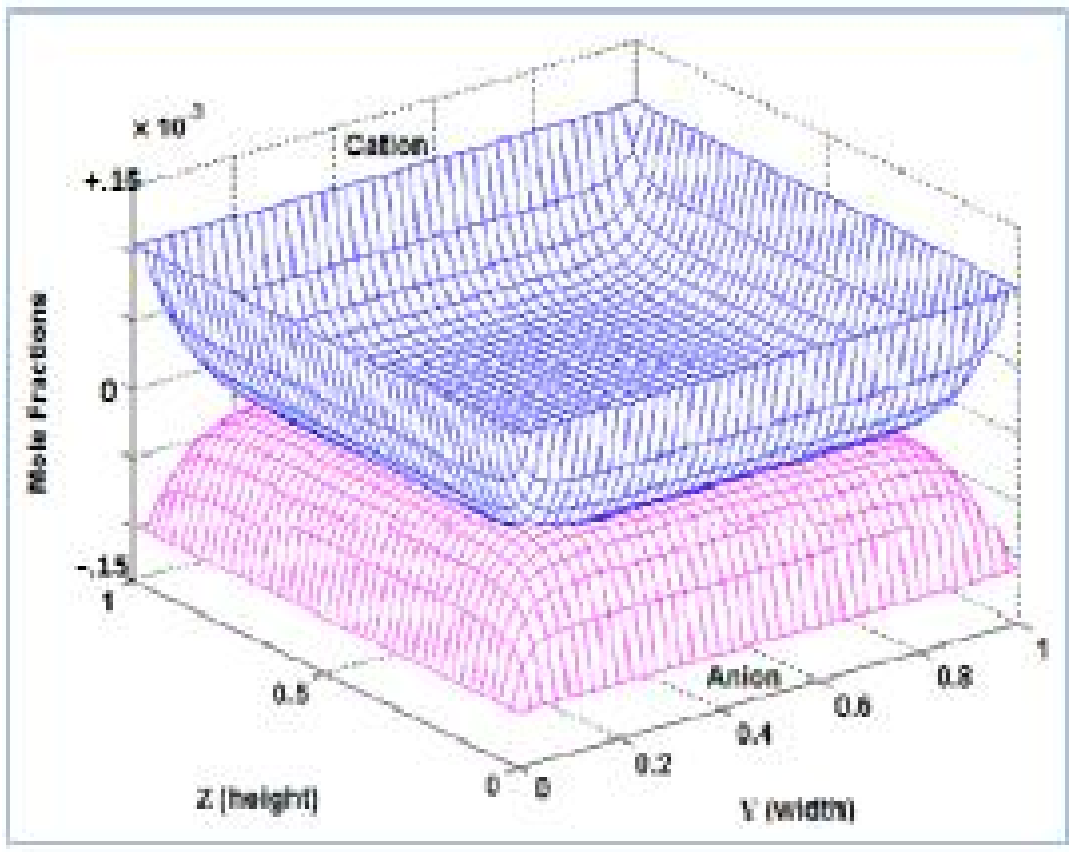

Figure 9. Dimensionless mole fraction distribution for an aqueous electrolyte solution in nanochannel 1 with height 20 nm and width 20 nm.

The dimensionless electrical potential and velocity variations in the nanochannel 2 are displayed in figure10. We observe that in the graph the central region is flat for channel 1, but records mount like formation for channel 2. This is due to the reduced dimensions of channel 2 both in width as well as height thus the distribution of the chosen parameters reaches the maximum value that lasts over a very small region, observed as the mount.

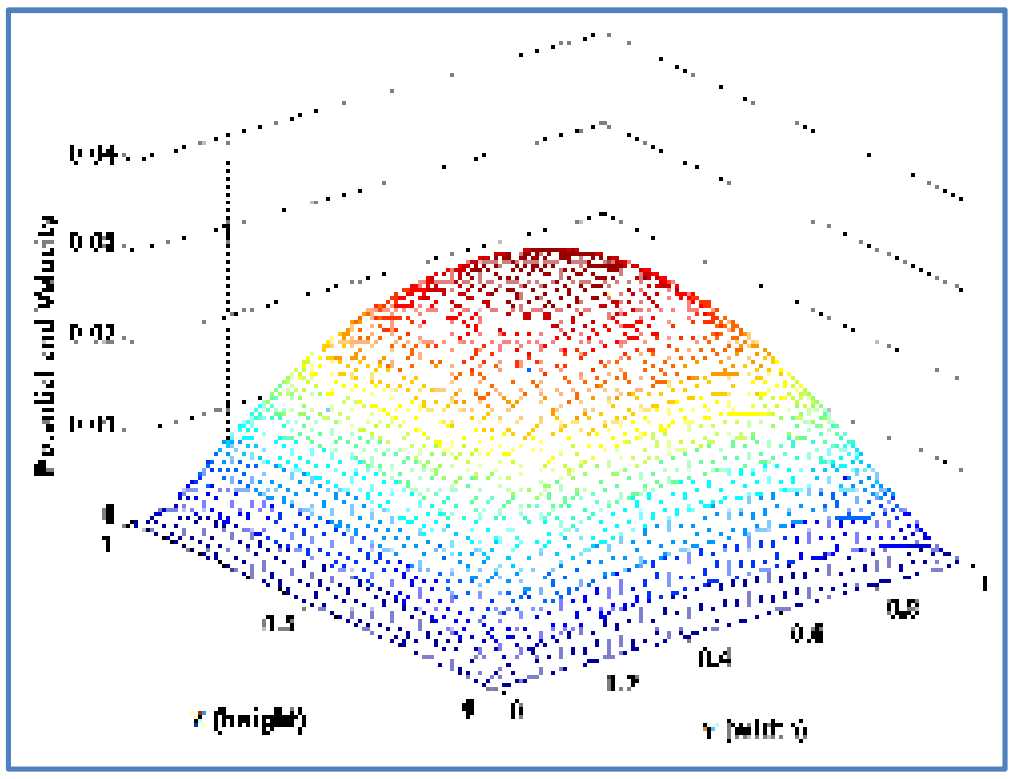

Figure 10. Dimensionless 'Ψ' and 'u' distribution for an aqueous electrolyte solution in nanochannel 2 with height 4 nm and width 4 nm.

The mole fractions of cation and anion species in nanochannel 2 are displayed in figure11. The values of Debye length and the height of the nanochannel 2 for this give $\varepsilon = 0.200$, hence the height of the channel is of the order of a fraction of the Debye length. This makes the height of the nanochannel 2 very small. Consequently, the cation and anion segments record a spatial separation showing the formation of two separate mounts, one of them appearing to be the mirror image of the other.

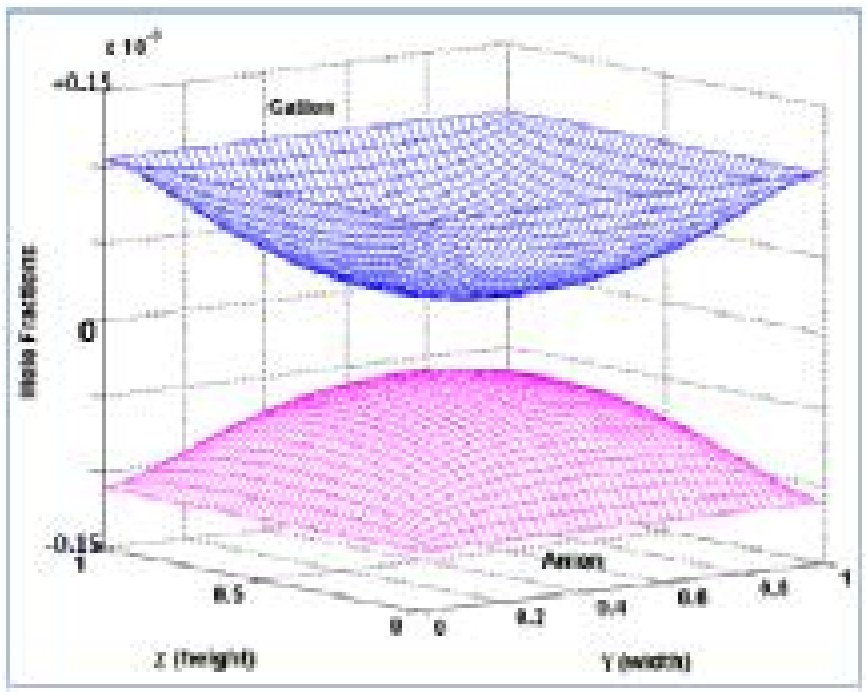

Figure 11. Dimensionless mole fraction distribution for an aqueous electrolyte solution in nanochannel 2 with height 4 nm and width 4 nm.

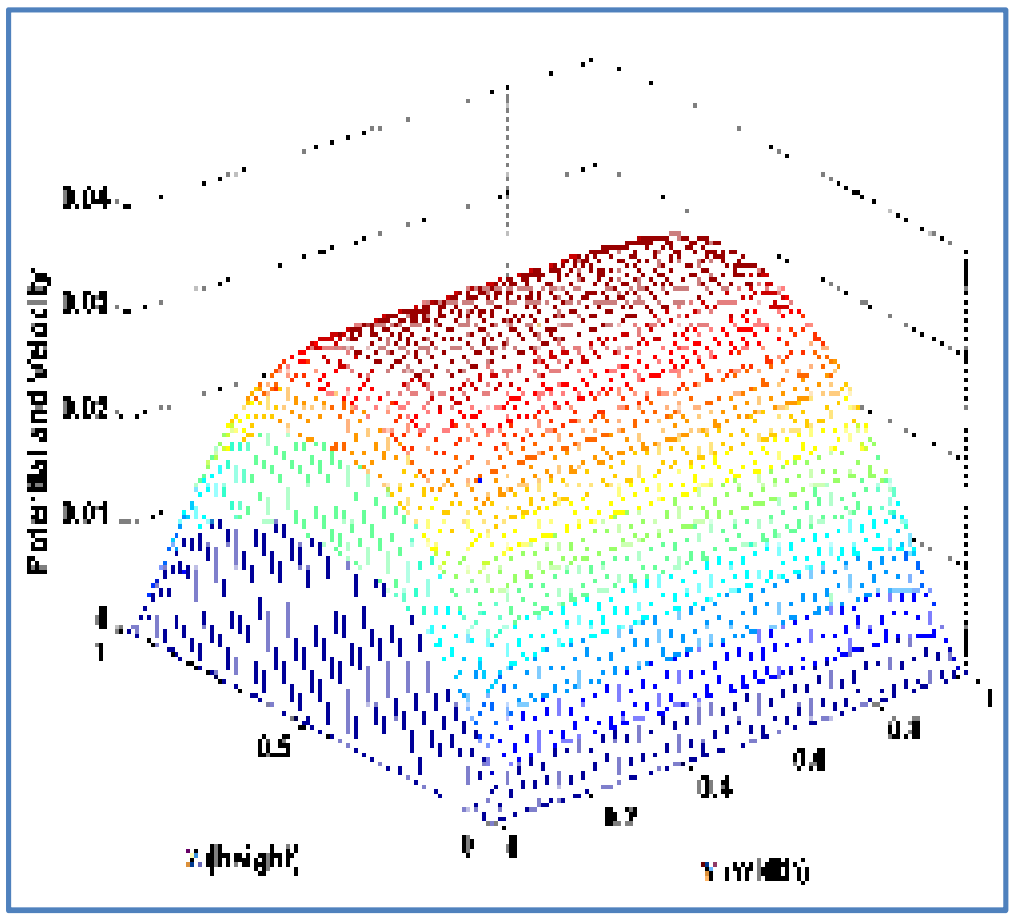

Figure 12. Dimensionless ‘Ψ’ and ‘u’ distribution for an aqueous electrolyte solution in nanochannel 3 with height 4 nm and width 20 nm.

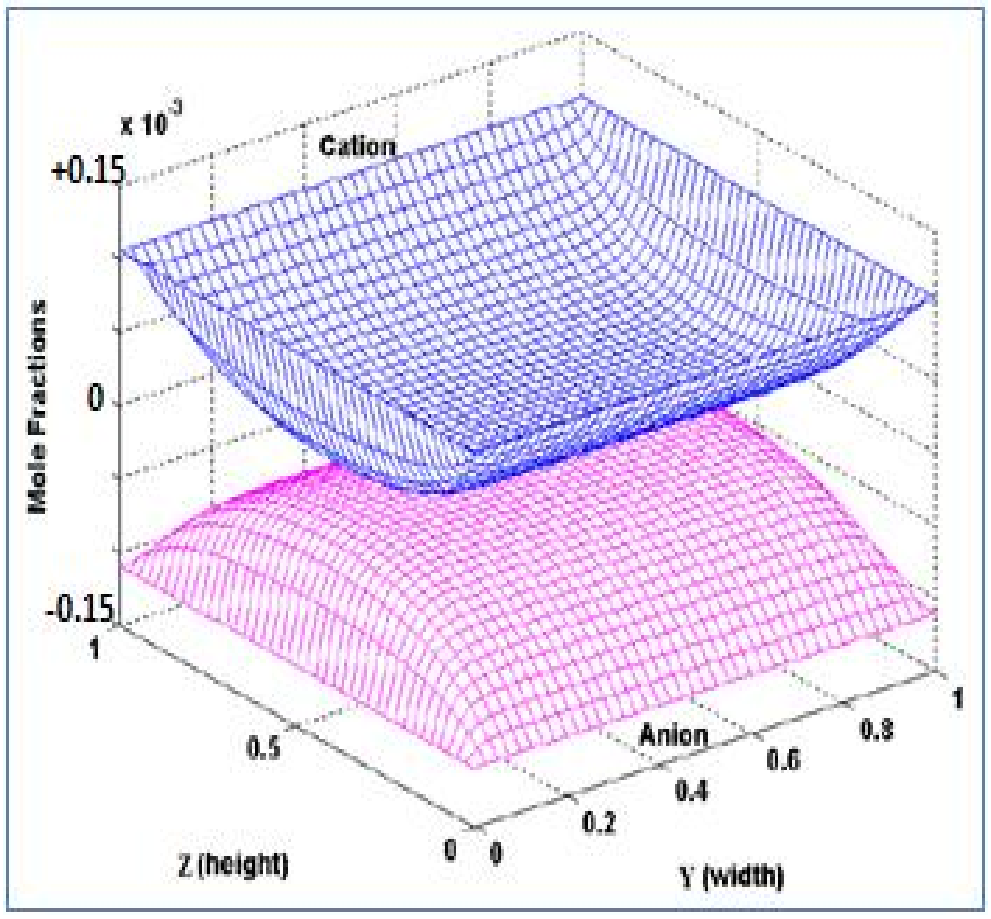

Figure 13. Dimensionless mole fraction distribution for an aqueous electrolyte solution in nanochannel 3 with height 4 nm and width 20 nm.

The electrical potential and velocity for nanochannel 3 are displayed in figure 12. In nanochannel 3, we choose the width = 20 nm, which is equals to width of the nanochannel 1 and the height = 4nm, which is equal to the height of nanochannel 2. The potential and velocity plots show similarity in the results that are predictable with the observations found with nanochannel 1 and 2. It is found that the mount formation about the height is continued in the result. Consequently, it indicates narrowing down of the height of the channel is responsible for this as was explained in the case of channel 2. However, the formation of the mount is not sharp; we find that the mount extends maintaining the same height about the width throughout. This concludes that the potential variation is uniform about the centre of the height of the channel 3.The mole fractions of cation and anion species in nanochannel 2 are displayed in figure 13 for mono-valent binary electrolyte in an aqueous solution.

## Conclusion

We applied the finite different technique to solve one and two dimensional coupled system of Navier – Stokes, Maxwell – Stefan and Poisson - Boltzmann equations, and the solutions give an extremely accurate prediction of the dielectric constant for a variety of salts and a wide range of concentrations. By incorporating the variation of dielectric constant of aqueous electrolyte solution into basic equations, improvement in the predictable nature of electrical potential, mass transport and velocity distribution in the electro-osmotic flow of a nano channel could be achieved.